\documentstyle[prc,aps,epsfig]{revtex}
\begin{document}
\draft
%@twocolumnfalse\endcsname

\title{\bf BEC and the particle mass}
\author{G.A.Kozlov}
\address{Bogoliubov Laboratory of Theoretical Physics,\\
Joint Institute for Nuclear Research, 141980 Dubna,
Moscow Region, Russia}

\date{\today}
\maketitle

\begin{abstract}
We derive new features of Bose-Einstein correlations
by means of Quantum Field Theory at finite temperature,
supplemented by operator-field evolution approach.
%of Langevin type, allows a deeper understanding of a coherent
%behaviour of the emitting source of elementary particles and
%an origin of the observed shape of two-particle BEC function.
The origin of the dependence of the measured two-particle
correlation function (as well as the so-called correlation radius)
on the particle mass has received its correct explanation. The
lower bound on the particle emitting source size is estimated.
\end{abstract}
%\pacs{PACS numbers: 25.75.Gz 12.40.Ee 03.65.-w 05.30.Jp}

\vspace{1 cm}

%{\bf 1}.
%{\centerline {\large{\bf 1. Introduction}}}
\section {Introduction}

Over the past few decades, a considerable number of studies have been
made on the phenomena of multiparticle correlations induced by collisions between
elementary particles and heavy ions. It is understood that the studies
of correlations between produced particles, the effects of coherence and chaoticity,
an estimation of particle emitting source size play the important roles
in high energy physics. Notice that produced particles affect the time evolution
of the system. What this does mean? The effect of quantum correlations between
particles is essentially a statistical mechanical phenomenon, and therefore cannot
be fully described only within quantum field theory (QFT) in closed form.
This problem becomes manifest when we define the basic operators in the QFT at
finite temperature, and when we derive the evolution of them. The common feature
is the coexistence of quantum and statistical fluctuations in one system composed particles.

In our previous papers [1,2], we established the master equation in the form
of the field-operator evolution (Langevin-like [3]) equation, which allows
one to gain a better understanding of  possible coherent behavior
of an emitting source of elementary particles.
The clear shapes of both two-particle Bose-Einstein
and Fermi-Dirac correlation functions were observed in the LEP experiments
ALEPH [4], DELPHI [5] and OPAL [6] which also indicated a dependence of
the measured correlation radius on the hadron ($\pi$, $K$ - mesons) mass. Recently,
the ZEUS
Collaboration at HERA reported [7] the results of  Bose-Einstein correlations (BEC)
between kaons (charged and neutral). The radii of production volume for neutral and
charged kaons turned out similar. There is no yet the definite explanation of
the origin of mass dependence above mentioned (see, e.g., [8]). Notice, that the LUND
model [9] does not predict such a dependence. One of the aims of this paper is to
clarify this question and to give an explanation of the origin of the
dependence of the radius of two-particle source emitter on a particle mass.

 We focused [2,1] on two specific
features of the Bose-Einstein correlations  clearly visible when
the latter are presented in the language of the QFT,
supplemented by the operator-field evolution.  The features we discussed were:\\
$(i)$ How the possible coherence of the hadronizing (or deconfined) system
(modeled here by the external
stationary force $P$ that appears in the Langevin-like equations) influences
the 2-particle BEC function
$C_2(Q=\sqrt {(p_{\mu} - p^{\prime}_{\mu})^{2}})$, in which two particles
are characterized by their four-momenta  $p_{\mu}$ and $p^{\prime}_{\mu}$ ? \\
$(ii)$ What is the true origin of the experimentally
observed $Q$-dependence of the $C_2(Q)$ in the approach used in [2,1]?

The only physical meaning of the memory term $\tilde K(p_{\mu})$, the noise
spectral function $\psi (p_{\mu})$, and the so-called
coherence function $\alpha = \alpha (m^2,\vec p^2,\beta)$ related to them
\begin{equation}
\alpha (m^2,\vec p^2,\beta)= \frac{M^{4}_{ch}\,\rho^{2}(\omega,\epsilon)\,P^2}
{{\vert\omega - \tilde{K}(p_{\mu})\vert}^{2}\,n(\omega,\beta)}
\,\,\, {\rm as}\,\, \epsilon > 0 \,\,\,\,{\rm and}\,\, Q^2 \rightarrow 0,
\label{e1}
\end{equation}
have not as yet been investigated carefully in [1].   In formula (\ref{e1}),
$\tilde{K}(p_{\mu}) =\psi (p_{\mu})\,\hat c(p_{\mu})$ with $[\hat c(p_{\mu}),
\hat c(p^{\prime}_{\mu})] = \delta{^4} (p_{\mu} -p^{\prime}_{\mu})$, where
$p_{\mu}=(\omega,\vec{p})$;
%$\omega$
%is the Fourier transformed component
%of 4-vector $p_{\mu}$ and it is conjugated to time $t$;
 $n(\omega,\beta) = \left\{\exp \left[(\omega - \mu)\beta\right] -
1\right\}^{-1}$ is the number of Bose-particles in the reservoir
characterized by the parameter $\mu$ (the chemical potential) and the inverse
temperature $\beta$.
We have already established [1] that there is at $p_{\mu} - p^{\prime}_{\mu}\neq 0$  a
finite volume $\Omega_{0}(r)$ for a two-particle emitter source, and there is a  correlation picture.
The space in which the massive fields settle has its own characteristic length
$L_{ch}\sim M^{-1}_{ch}$, in the same sense that the massive field with mass $m_{c}$ for
the Compton length $\lambda_{c} = m^{-1}_{c}$ is served. The appearance of the $\delta$-like
distribution
$\rho(\omega,\epsilon)= \int dx\,e^{(i\omega -\epsilon)x}\rightarrow \delta (\omega)$ as
$\epsilon\rightarrow 0$ is a consequence of the finite volume of produced particles
(formally, $\epsilon$ is given by the physical spectrum).
The volume $\Omega_{0}(r)$ is conjugated to
a characteristic mass scale $M_{ch}$, $\Omega_{0}(r)\sim 1/{M^{4}_{ch}}$ in such a way that
$M_{ch}\rightarrow\infty$ when the correlation domain
does not exist. Hence, $M_{ch}\rightarrow 0$ obeys the condition  $\alpha\rightarrow 0$,
which means there is no distortion (under the constant force $P$) acting on the system of produced particles.
On the other hand,
any distortion that disturbs this system for any reason (the coherence function $\alpha > 0$)
leads to a finite
domain of produced particles ($M_{ch}\neq 0$). Rather, a strong
distortion, $\alpha\rightarrow\infty$, satisfies the point-like region of the particle
source emitter.

It is difficult to derive both
$\tilde K(p_{\mu})$ and $\psi (p_{\mu})$ using the general properties
of QFT. It is worth stressing  that knowledge of one of these functions leads
to an understanding of the other, due of the necessary condition [2]
%($p_{\mu}=(p^{0}=\omega,\vec p)$)
\begin{equation}
\int^{+\infty}_{-\infty}\frac{d\omega}{2\,\pi}{\left\vert
\frac{\psi (p_{\mu})}{\omega - \tilde K(p_{\mu})}\right\vert}^{2} = 1,
\label{e2}
\end{equation}

which is nothing other than the generalized fluctuation-dissipation theorem.
The spectral properties
%\begin{equation}
$$ {\vert\psi (p_{\mu})\vert}^2\rightarrow\frac{2}{\omega}\,\sin(\beta\omega)
{\vert\omega - \tilde K(p_{\mu})\vert}^2,\,\,\, 0<\beta <\infty $$
%\label{e3}
%\end{equation}
 follow from Eq. (\ref{e2}) with
\begin{equation}
 \lim_{\beta\rightarrow\infty} \frac{1}{\pi}\,\frac{\sin(\beta\omega)}{\omega}
= \delta(\omega).
\label{e4}
\end{equation}

%{\bf 2}.
%{\centerline {\bf {2. Green's function}}}
\section { Green's function}

In this paper, we would like to focus on the role of the particle mass, which
 influences the correlations between particles. To solve this problem,
one must derive the memory term $\tilde K(p_{\mu})$ using the general properties
of QFT.

We suppose that we are working with fields that correspond to a thermal
field $\Phi(x)$ with the standard definition of the Fourier transformed propagator
$F[\tilde G(p)]$
\begin{equation}
F[\tilde G(p)]= G(x-y) = Tr\left\{T[\Phi(x)\Phi(y)]\rho_{\beta}\right\},
\label{e5}
\end{equation}
with $\rho_{\beta}= e^{-\beta H}/Tr e^{-\beta H}$ being the density matrix of a local
system in equilibrium at temperature $T=\beta^{-1}$ under the Hamiltonian $H$.

We consider the interaction of $\Phi(x)$ with the external scalar field
given by the potential $U$. In contrast to an electromagnetic field, this potential
is a scalar one, but it is not a component of the four-vector. The Lagrangian
density can be written
%\begin{equation}
$$ L(x) = \partial_{\mu}\Phi^{\star}(x)\partial^{\mu}\Phi(x) -
(m^2 + U)\Phi^{\star}(x)\Phi(x) $$
%\label{e6}
%\end{equation}
and the equation of motion is
\begin{equation}
(\nabla^2 + m^2)\Phi(x) = -J(x),
\label{e7}
\end{equation}
where $J(x) = U\Phi(x)$ is the source density operator. A simple model like this allows one
to investigate the origin of the
%occurrence of the condensate in the static restricted
%potential of the external field. We are interested in the origin of the
unstable state
of the thermalized equilibrium in a nonhomogeneous external field under the influence
of  source density operator $J(x) = U\Phi(x)$. For example, the source can be considered
as $\delta$-like generalized function, $J(x)=\tilde\mu\,\rho(x,\epsilon)\Phi(x)$, in which
$\rho(x,\epsilon)$ is a $\delta$-like succession giving the $\delta$-function as
$\epsilon\rightarrow 0$ (where $\tilde\mu$ is some massive parameter). This model is useful
because the $\delta$-like potential $U(x)$ provides the model conditions
for  restricting the particle emission domain (or the deconfinement region). We suggest the
following form:
%\begin{equation}
$$ J(x) = - \Sigma(i\partial_{\mu})\,\Phi(x) + J_{R}(x), $$
%\label{e8}
%\end{equation}
where the source $J(x)$ decomposes into a regular systematic motion part
$\Sigma(i\partial_{\mu})\,\Phi(x)$ and the random source $J_{R}(x)$. Thus, the
equation of motion (\ref{e7}) becomes
%\begin{equation}
$$ [\nabla^2 + m^2 - \Sigma(i\partial_{\mu})]\Phi(x) = -J_{R}(x), $$
%\label{e9}
%\end{equation}
and the propagator satisfies the following equation:
\begin{equation}
[-p^{2}_{\mu} +m^2 -\tilde\Sigma(p_{\mu})]\tilde G(p_{\mu}) =1.
\label{e10}
\end{equation}
The random noise is introduced with a random operator
$\eta(x) = - m^{-2}\,\Sigma(i\partial_{\mu})$, for that the equation of motion
looks like:
\begin{equation}
 \{\nabla^2 + m^2[1+\eta (x)]\}\Phi(x) = -J_{R}(x).
\label{e9}
\end{equation}

We assume that $\eta (x)$  varies stochastically with the certain  correlation function
(CF), e.g.,  the Gaussian CF
$$\langle\eta(x)\,\eta(y)\rangle = C\exp(-z^{2}\mu^{2}_{ch}),\,\,\, z= x-y ,$$
where $C$ is the strength of the noise described by the distribution function
$\exp(-z^{2}/L^{2}_{ch})$ with  $L_{ch}$ being the noise characteristic
scale. Both $C$ and $\mu_{ch}$ define the influence of the (Gaussian)
noise on the correlations between particles that "feel" an action of an environment.
The solution of Eq. (\ref{e9}) is
\begin{equation}
\Phi(x) = -\int dy\, G (x,y)\,J_{R}(y),
\label{e99}
\end{equation}
where the Green's function obeys the Eq.
$$\{\nabla^2 + m^2[1+\eta (x)]\}G(x,y) = \delta(x-y).$$

The final aim might having been to find the solution of Eq. (\ref{e99}), and then
average it over random operator $\eta (x)$. Note that the operator
$M(x) =\nabla^2 + m^2[1+\eta (x)]$ in the causal Green's function
$$G(x,y) = \frac{1}{M(x) +i\,o}\delta (x-y)$$ is not definitely positive.
However, we shall formulate another approach, where the random force influence
are introduced on the particle operator level.

We now  introduce the general non-Fock representation
%of the canonical commutation
%relation (CCR). To this, we consider the
in the form of the operator generalized functions
\begin{equation}
b(x) = a(x) + R(x),
\label{e11}
\end{equation}
\begin{equation}
b^{+}(x) = a^{+}(x) + R^{+}(x),
\label{e12}
\end{equation}
where the operators $a(x)$ and $a^{+}(x)$ obey the canonical commutation
relations (CCR):
%\begin{equation}
$$ [a(x),a(x^{\prime})]= [a^{+}(x),a^{+}(x^{\prime})]= 0, $$
%\label{e13}
%\end{equation}
%\begin{equation}
$$ [a(x),a^{+}(x^{\prime})]= \delta (x-x^{\prime}). $$
%\label{e14}
%\end{equation}
The operator-generalized functions $R(x)$ and $R^{+}(x)$ in (\ref{e11}) and
(\ref{e12}), respectively, include random features
describing the action of the external forces.

Both $b^{+}$ and $b$ obviously define the CCR representation. For each function
$f$ from the space $S(\Re_{\infty})$ of smooth decreasing functions, one can
establish new operators $b(f)$ and $b^{+}(f)$
%\begin{equation}
$$ b(f) = \int f(x) b(x) dx = a(f) +\int f(x) R(x) dx, $$
%\label{e15}
%\end{equation}
%\begin{equation}
$$ b^{+}(f) = \int \bar f(x) b^{+}(x) dx = a^{+}(f) +\int\bar f(x) R^{+}(x) dx. $$
%\label{e16}
%\end{equation}
The transition from the operators $a(x)$ and $a^{+}(x)$ to $b(x)$ and $b^{+}(x)$,
obeying those commutation relations as $a(x)$ and $a^{+}(x)$, leads
to linear canonical representations. If both $b(f)$ and $b^{+}(f)$ create the
Fock representation of the CCR, one can then find the operator $\hat U$ that obeys the
following conditions:
%\begin{equation}
$$ \hat U a(f) \hat U^{-1} = b(f), $$
%\label{e17}
%\end{equation}
%\begin{equation}
$$ \hat U a^{+}(f) \hat U^{-1} = b^{+}(f). $$
%\label{e18}
%\end{equation}
We now move to a simple  physical pattern.
Let us define the differential evolution (in time) equation, where the
sharp and chaotically fluctuating function obeying this equation
is the main object. Since we are dealing with continuous time, we can
formulate the stochastic differential equation applied to each
(analytical) function under the distortion of a random force. In classical
mechanics, the stochastic processes in a dynamic system are under the
weak action of a "large" system [10]. "Small" and "large" systems
are understood to mean that the number of the states of freedom of
the former is less than that of the latter. We do not exclude even
the possibility of interplay between these systems. When the "large" system
is in an equilibrium state (e.g., a thermostat state), our method
allows one to describe the approximation to the distribution of the
probability in order to find the physical states in the "small" system. On the quantum
level, the role of such a "small" system is played by the restricted region
of produced particles, a particle source emitter with a definite size,
that we study in this paper.

%{\bf 3. }
%\centerline {\bf {3. Evolution equation}}}
\section {Evolution equation}

Following the idea of classical
Brownian motion [11,3] of a particle with a unit mass, a charge $g$, and velocity $v(t)$
in the external, let us say, electric field $E$, one can write in the following form
the formal equation describing the evolution (in real time $t$) of this particle:
%\begin{equation}
$$ \partial_{t}v(t) = \gamma\,v(t) + F + g\,E, $$
%\label{e19}
%\end{equation}
where $F$ stands for a random force subject to the Gaussian white noise, and
$\gamma$ is a friction coefficient.

%In what concerns the first point we have obtained identical
%expression for $C_2(Q)$ as derived in \cite{Weiner} by means of a quantum
%statistical (QS) approach with novel interpretation of the {\it
%chaoticity} parameter $p$ introduced there.

%In the second point we demonstrate that in order to obtain a
%given (experimentally observed) shape of $C_2(Q)$ (i.e., its $Q$-dependence, where
%$Q=|k_{\mu}-k'_{\mu}| = \sqrt{(k_{\mu}- k'_{\mu})^2}$) one has to
%account somehow for  the {\it finiteness} of the space-time
%region of the particle emission (i.e., of a particle production {\it
%source}). In our approach this means the necessity of smearing
%out of delta functions $\delta(Q_{\mu})$ appearing in the definition of
%thermal averages of some operators, which is characteristic feature
%of QFT approach used here.
%The freedom in using different types of
%smearing functions to perform such a procedure allows us to account
%for all possible different shapes of hadronizing sources apparently
%observed by experiment.

Referring to [2,1] for details, let us recapitulate here the
main points of our approach in the quantum case: the collision process
produces a number of
particles, out of which we select only one (we assume for simplicity that
we are dealing only with identical bosons) and describe it by stochastic
operators $b(\vec{p},t)$ and $b^{+}(\vec{p},t)$, carrying the features of
annihilation and creation operators, respectively.
The rest of the particles
are then assumed to form a kind of heat bath, which remains in
an equilibrium characterized by a temperature $T$ (one of our parameters).
% All averages $\langle (\dots) \rangle$ are
%therefore thermal averages of the type: $\langle (\dots) \rangle
%=Tr\left[(\dots) e^{-\beta H}\right]/Tr\left( e^{-\beta H}\right)$.
We also allow for some external (relative to the above heat bath)
influence on our system.
%Therefore we shall represent the operator
%$b(\vec{k},t)$ as consisting of a part corresponding to the action
%bel{eq:apR}
%\end{equation}
The time evolution of such a system is then assumed to be given by a
Langevin-type equation [2,1] for the new stochastic operator $b(\vec{p},t)$
\begin{equation}
i\partial_t b(\vec{p},t) =  A(\vec{p},t) + F(\vec{p},t) + P
\label{e20}
\end{equation}
(and a similar conjugate equation for $b^{+}(\vec{p},t)$). We assume
an asymptotic free undistorted operator  $a(\vec{p},t)$, and that the deviation
from the asymptotic free state is provided by the random operator
$R(\vec{p},t)$: $a(\vec{p},t)\rightarrow b(\vec{p},t) = a(\vec{p},t) +
R(\vec{p},t)$. This means, e.g., that the particle density number
(a physical number) ${\langle n(\vec{p},t)\rangle}_{ph} =
\langle n(\vec{p})\rangle + O (\epsilon)$, where ${\langle n(\vec{p},t)\rangle}_{ph}$
means the expectation value of a physical state, while ${\langle n(\vec{p})\rangle}$
denotes that of an asymptotic state. If we ignore the deviation
from the asymptotic state in equilibrium, we obtain an ideal fluid.
One otherwise has to consider the dissipation term;  this is why
 we use the Langevin scheme to derive the evolution equation,
 but only on the quantum level. We derive the evolution equation in an integral
form that reveals the effects of thermalization.

 Equation (\ref{e20}) is supposed to model all aspects of the hadronization processes
(or deconfinement). The combination $A(\vec{p},t)+ F(\vec{p},t)$ in the r.h.s of
(\ref{e20}) represents the
so-called {\it Langevin force} and is therefore responsible for the
internal dynamics of particle emission, as the memory term $A$ causes
dissipation and is related to stochastic dissipative forces [2]
%\begin{equation}
$$ A(\vec{p},t) = \int^{+\infty}_{-\infty}\! d\tau K(\vec{p},t-\tau)
b(\vec{p},\tau) $$
%\label{e21}
%\end{equation}
with  $K(\vec{p},t)$ being the kernel operator describing the
virtual transitions from one (particle) mode to another.
At any dependence of the field operator $K$ on the time, the function
$A(\vec{p},t)$ is defined by the behavior of the system at the precedent moments.
The operator $F(\vec{p},t)$ is responsible for the action of
a heat bath of absolute temperature $T$ on a particle in the heat bath, and
under the appropriate circumstances is given by
%\begin{equation}
$$ F(\vec{p},t) =
\int^{+\infty}_{-\infty}\!\frac{d\omega}{2\pi}\psi(p_{\mu})\hat{c}(p_{\mu})
e^{-i\omega t} . $$
%\label{e22}
%\end{equation}
The heat bath is represented by an ensemble of coupled oscillators,
each described by the operator $\hat{c}(p_{\mu})$ such
that $\left[\hat{c}(p_{\mu}),\hat{c}^{+}(p'_{\mu})\right] =
\delta^4(p_{\mu}-p'_{\mu})$, and is characterized by the noise spectral function
$\psi(p_{\mu})$ [2,1]. Here, the only statistical assumption is that the heat bath
is canonically distributed. The oscillators are coupled to a particle, which is
in turn acted upon by an outside force.
Finally, the constant term $P$ in (\ref{e20}) (representing {\it an external source}
term in the Langevin equation) denotes a possible influence of
some external force. This force
would result, e.g., in a strong ordering of phases leading
therefore to a coherence effect.

The solution of equation (\ref{e20}) is given in $S(\Re_{4})$ by
\begin{equation}
\tilde b(p_{\mu}) = \frac{1}{\omega - \tilde K(p_{\mu})}\,
[\tilde F(p_{\mu}) + \rho (\omega_{P},\epsilon)],
\label{e23}
\end{equation}
where $\omega$ in $\rho (\omega,\epsilon)$ was replaced by  new scale
$\omega_{P} = \omega/P$.
It should be stressed that the term containing
$\rho (\omega_{P},\epsilon)$ as $ \epsilon\rightarrow 0$ yields the general solution to
Eq. (\ref{e20}). Notice that the distribution $\rho (\omega_{P},\epsilon)$  indicates
the continuous character of the spectrum, while the arbitrary small quantity
$\epsilon$ can be defined by the special physical conditions or the physical
spectra. On the other hand, this $\rho (\omega_{P},\epsilon)$ can be understood as
temperature-dependent succession (\ref{e4}), in which  $\epsilon\rightarrow \beta^{-1}$.
Such a succession yields the restriction on the $\beta$-dependent
second term in the solution (\ref{e23}), where at small enough $T$ there is
a narrow peak at $\omega = 0$.

 From the scattering matrix point of view, the solution (\ref{e23}) has the following
physical meaning: at a sufficiently  outgoing past and future, the fields described by the
operators $\tilde a(p_{\mu})$ are free and,  the initial and the final states
of the dynamic system are thus characterized by  constant amplitudes.
Both  states, $\varphi (-\infty)$ and $\varphi (+\infty)$, are related to
one another by an operator $S(\tilde R)$ that transforms  state
$\varphi (-\infty)$ to  state $\varphi (+\infty)$ while depending on the behaviour
of $\tilde R(p_{\mu})$:
$$\varphi (+\infty) = \varphi (\tilde R) = S(\tilde R)\varphi (-\infty).$$
In accordance with this definition, it is natural to identify $S(\tilde R)$ as the
scattering matrix in the case of arbitrary sources that give rise to the intensity of
$\tilde R$.

Based on QFT point of view, relation (\ref{e11}) indicates the
appearance of the terms containing nonquantum fields that are characterized by
the operators $\tilde R(p_{\mu})$.
Hence, there are terms with $\tilde R$ in the matrix elements, and
these $\tilde R$ cannot be realized via real particles. The operator function
$\tilde R(p_{\mu})$ could be considered as the limit on an average value of some quantum
operator (or even a set of operators) with an intensity that increases to infinity.
The later statement can be visualized in the following mathematical representation [2]:
$$\tilde R(p_{\mu}) = \sqrt {\alpha\,\Xi (p_{\mu},p_{\mu})},\,\,
\Xi (p_{\mu},p_{\mu}) = {\langle\tilde a^{+}(p_{\mu})\, \tilde a(p_{\mu})\rangle}_{\beta} ,
 $$
where $\alpha$ is the coherence function that gives the strength of the average
$\Xi (p_{\mu},p_{\mu})$.
%We shall find this coherence function carrying the
%dependence on $\beta$, the particle mass $m$ and the chemical potential $\mu$.

In principal, interaction with the fields described by $\tilde R$ is provided by
the virtual particles, the propagation process of which is given by the potentials
defined by the $\tilde R$ operator function.

The condition $M_{ch}\rightarrow 0$ (or $\Omega_{0}(r)\sim\frac{1}{M^{4}_{ch}}\rightarrow\infty$)
in the representation
$$\lim_{p_{\mu}\rightarrow p^{\prime}_{\mu}}\Xi (p_{\mu},p^{\prime}_{\mu}) =
\lim_{Q^{2}\rightarrow 0} \Omega_{0}(r)\,n(\bar\omega,\beta)\exp (-q^{2}/2)\rightarrow
\frac{1}{M^{4}_{ch}}\,n(\omega,\beta), $$
with [2]
$$ \Omega_{0}(r)=\frac{1}{\pi^2}\,r_{0}\,r_{z}\,r^{2}_{t}
 $$
means that the role of the arbitrary source characterized by the operator function
$\tilde R(p_{\mu})$ in $\tilde b(p_{\mu})= \tilde a(p_{\mu}) + \tilde R(p_{\mu})$ disappears.

%{\bf 4.}
%{\centerline {\bf {4. Green's function and kernel operator}}}
\section{ Green's function and kernel operator}

Let us go to the thermal field operator $\Phi(x)$ by means of the linear combination
of the frequency parts $\phi^{+}(x)$ and $\phi^{-}(x)$
\begin{equation}
\Phi (x) = \frac{1}{\sqrt{2}}\,\left [\phi^{+}(x) +\phi^{-}(x)\right ]
\label{e24}
\end{equation}
composed of the operators $\tilde b(p_{\mu})$ and $\tilde b^{+}(p_{\mu})$ as the
solutions of equation (\ref{e20}) and conjugate to it, respectively:
%\begin{equation}
$$ \phi^{-} (x) = \int \frac{d^{3}\vec{p}}{(2\pi)^{3} 2 (\vec p^{2} +m^{2})^{1/2}}
\tilde b^{+}(p_{\mu})\,e^{ipx}, $$
%\label{e25}
%\end{equation}
%\begin{equation}
$$\phi^{+} (x) = \int \frac{d^{3}\vec{p}}{(2\pi)^{3} 2 (\vec p^{2} +m^{2})^{1/2}}
\tilde b(p_{\mu})\,e^{-ipx}. $$
%\label{e26}
%\end{equation}

One can easily find two equations of motion for the Fourier transformed operators
$\tilde b(p_{\mu})$ and $\tilde b^{+}(p_{\mu})$ in $S(\Re_{4})$
\begin{equation}
[\omega - \tilde K(p_{\mu})]\tilde b(p_{\mu}) = \tilde F(p_{\mu}) + \rho(\omega_{P},\epsilon),
\label{e27}
\end{equation}
\begin{equation}
[\omega - \tilde K^{+}(p_{\mu})]\tilde b^{+}(p_{\mu}) =
\tilde F^{+}(p_{\mu}) + \rho(\omega_{P},\epsilon),
\label{e28}
\end{equation}
which are transformed into new equations for the frequency parts
$\phi^{+} (x)$ and $\phi^{-} (x)$ of the field operator $\Phi (x)$ (\ref{e24})
\begin{equation}
i\partial_{0}\phi^{+} (x) - \int_{\Re_{4}} K(x-y)\,\phi^{+} (y)dy =
F(x) + P\,\partial_{0}D(x)\,e^{-\epsilon t},
\label{e29}
\end{equation}
\begin{equation}
- i\partial_{0}\phi^{-} (x) - \int_{\Re_{4}} K^{+}(x-y)\,\phi^{-} (y)dy =
F^{+}(x) + P\,\partial_{0}D(x)\,e^{-\epsilon t}.
\label{e30}
\end{equation}
Here, the field components $\phi^{+}(x)$ and  $\phi^{-}(x)$ are nonlocalized under the effect
of the invariant formfactors $K(x-y)$ and $K^{+}(x-y)$, respectively. In general,
these formfactors can
admit the description of locality for nonlocal interactions. The function $D(x)$  in
Eqs. (\ref{e29}) and (\ref{e30}) obeys the commutation relation
%\begin{equation}
$$[\Phi (x),\Phi (y)]_{-} = - i D(x) $$
%\label{e31}
%\end{equation}
and can be written [12]
%\begin{equation}
$$ D(x)= \frac{1}{2\,\pi}\,\epsilon(x^{0})\,\left [\delta(x^2) -
\frac{m}{2\,\sqrt{x^{2}_{\mu}}}\,\Theta(x^2)\,J_{1}\left (m\sqrt{x^2_{\mu}}\right )\right ], $$
%\label{e32}
%\end{equation}
where $\epsilon(x^{0})$ and $\Theta(x^2)$ are the standard unit and the step functions,
respectively, while $ J_{1}(x)$ is the Bessel function. On the mass-shell,
$D(x)$ becomes
%\begin{equation}
$$ D(x)\simeq \frac{1}{2\,\pi}\,\epsilon(x^{0})\,\left [\delta(x^2) -
\frac{m^2}{4}\,\Theta(x^2) \right ]. $$
%\label{e33}
%\end{equation}
At this stage, it must be stressed that we have new generalized evolution
Eqs. (\ref{e29}) and (\ref{e30}), which retain  the general
features of the propagating and
interacting of the quantum fields with mass $m$ that are in the heat bath (reservoir)
and are chaotically distorted by  other fields. For  further analysis,
let us rewrite the system of Eqs. (\ref{e29}) and (\ref{e30}) in the following form:
\begin{equation}
i\partial_{0}\phi^{+} (x) -  K(x)\star\phi^{+} (x) = f(x),
\label{e34}
\end{equation}
\begin{equation}
- i\partial_{0}\phi^{-} (x) -  K^{+}(x)\star\phi^{-} (x) = f^{+}(x),
\label{e35}
\end{equation}
where $A(x)\star B(x)$ is the convoluted function of the generalized functions
$A(x)$ and $B(x)$, and
%\begin{equation}
$$ f(x) = F(x) + P\,\partial_{0}D(x)e^{-\epsilon t}. $$
%\label{e36}
%\end{equation}
Applying the direct Fourier transformation to both sides of Eqs.
(\ref{e34}) and (\ref{e35}) with the following properties of the
Fourier transformation
%\begin{equation}
$$ F[K(x)\star\phi^{+} (x) ] = F[K(x)]F[\phi^{+} (x)],$$
%\label{e37}
%\end{equation}
we  get two equations
\begin{equation}
[- p^{0} - \tilde K(p_{\mu})]\tilde\phi^{+}(p_{\mu}) = F[f(x)],
\label{e38}
\end{equation}
\begin{equation}
[p^{0} - \tilde K^{+}(p_{\mu})]\tilde\phi^{-}(p_{\mu}) = F[f^{+}(x)].
\label{e39}
\end{equation}
Multiplying Eqs. (\ref{e38}) and (\ref{e39}) by $p^{0} - \tilde K^{+}(p_{\mu})$
and $- p^{0} - \tilde K(p_{\mu})$, respectively, we find
\begin{equation}
[- p^{0} - \tilde K(p_{\mu})][p^{0} - \tilde K^{+}(p_{\mu})]\tilde\Phi(p_{\mu}) = T(p_{\mu}),
\label{e40}
\end{equation}
where
%\begin{equation}
$$ T(p_{\mu}) = [p^{0} - \tilde K^{+}(p_{\mu})]F[f(x)]-
 [p^{0} + \tilde K(p_{\mu})]F[f^{+}(x)]. $$
%\label{e41}
%\end{equation}
We are now  at the stage of the main strategy:  we have to identify the field
$\Phi (x)$ introduced in Eq. (\ref{e5}) and the field $\Phi (x)$ (\ref{e24}) built up of the
fields $\phi^{+}$ and $\phi^{-}$ as the solutions of  generalized
Eqs. (\ref{e29}) and (\ref{e30}). The next step is our requirement that
Green's function
$\tilde G(p_{\mu})$ in Eq. (\ref{e10}) and  the function $ \Gamma(p_{\mu},\beta)$,
that satisfies Eq.(\ref{e40})
\begin{equation}
[p^{0} + \tilde K(p_{\mu})][- p^{0} + \tilde K^{+}(p_{\mu})]\tilde\Gamma(p_{\mu}) = 1,
\label{e42}
\end{equation}
must be equal to each other, where [12]
%\begin{equation}
$$ \tilde G(p_{\mu})\rightarrow \tilde G(p^2, g^2, m^2)\simeq \frac{1 - g^2\,\xi(p^2, m^2)}
{m^2 - p^2 -i\epsilon} $$
%\label{e43}
%\end{equation}
with $g$ being the scalar coupling constant and the one-loop correction of the scalar field
$\xi << 1/m^2$ at $1/4 \leq (m^2/p^2)\leq 1$.
This means  we define the operator kernel $\tilde K(p_{\mu})$ in (\ref{e27}) from
the condition of the nonlocal coincidence of the Green's function $\tilde G(p_{\mu})$
in Eq. (\ref{e10}), and the thermodynamic function $\tilde\Gamma(p_{\mu},\beta)$
from (\ref{e42}) in $S(\Re_{4})$
$$ \lim_{x-x^{\prime}\sim O(r)} F[\tilde G(p_{\mu}) - \tilde\Gamma(p_{\mu},\beta)] =0.$$

We can easily derive the kernel operator
$\tilde K(p_{\mu})$ in the form
\begin{equation}
\tilde K(p_{\mu}) = {(m^2 + \vec p^2)}^{\frac{1}{2}}{\left [ 1 + g^2\xi(p^2, m^2)
\left ( 1- \frac{\omega^2}{m^2 + \vec p^2}\right )\right ]}^{\frac{1}{2}},
\label{e44}
\end{equation}
where
%\begin{equation}
$$ \xi(m^2) = \frac{1}{96\,\pi^2\,m^2}\left (\frac{2\,\pi}{\sqrt{3}} -1\right ),
\,\,\, p^2\simeq m^2, $$
%\label{e45}
%\end{equation}
and
%\begin{equation}
$$ \xi(p^2, m^2) = \frac{1}{96\,\pi\,m^2}\left (i\, \sqrt {1-\frac{4\,m^2}{p^2}}
+ \frac{\pi}{\sqrt{3}}\right ),
\,\,\, p^2\simeq 4 m^2. $$
%\label{e46}
%\end{equation}
The ultraviolet behaviour at $ \vert p^2\vert >> m^2$ leads to

%\begin{equation}
$$ \xi(p^2, m^2) \simeq  \frac{-1}{32\,\pi^{2}\,p^2}\left [\ln\frac{\vert p^2\vert}{m^2} -
\frac{\pi}{\sqrt{3}} - i\,\pi\Theta (p^2)\right ]. $$
%\label{e47}
%\end{equation}

%{\bf 5.}
%{\centerline {\bf {5. Correlation function }}}
\section { Correlation function }

Out of many details (for which we refer to [2,1])
%important in our case is the fact that
the $2$-particle BEC function
for identical particles is defined as
\begin{eqnarray}
C_2(Q) &=&  \chi(N) \cdot \frac{\tilde{f}(p_{\mu},p'_{\mu})}
{\tilde{f}(p_{\mu})\cdot\tilde{f}(p'_{\mu})}\,\Delta (r_{f}Q)
\simeq \chi(N) \cdot \left[1\, +\, D(p_{\mu},p'_{\mu})\right]\,
(1+ \sum_{l\geq 1}(r_{f}\cdot Q)^{l}),
\label{e48}
\end{eqnarray}
where $\tilde{f}(p_{\mu},p'_{\mu}) = \langle
\tilde{b}^{+}(p_{\mu})\tilde{b}^{+}(p'_{\mu})
\tilde{b}(p_{\mu})\tilde{b}(p'_{\mu})\rangle$ and
$\tilde{f}(p_{\mu}) = \langle
\tilde{b}^{+}(p_{\mu})\tilde{b}(p_{\mu})\rangle$ are the
corresponding thermal statistical averages with
$\tilde{b}(p_{\mu})$
being the corresponding Fourier transformed solution (\ref{e23}).
The multiplicity $N$ depending factor is equal to
$\chi (N) = \langle N(N-1)\rangle/\langle N\rangle^2$.
We have introduced in (\ref{e48}) the factor
$\Delta (r_{f}Q)\simeq 1+ \sum_{l\geq 1}(r_{f}\cdot Q)^{l}$,
which is nothing other than the consequence of the Bogolyubov's principle
of weakening of correlations at large distances
%which is
%called
%"adiabatic correlation-fluctuation factor"
that is characterized by the parameter of weakening of correlations
$r_{f}$ relevant to the particle
pair correlations. In the limit of infinitely small scale $r_{f}$
the function $C_2(Q)$ coincides with that obtained in [2]. The latter
statement  means that one returns to the ideal correlation pattern.

As was shown in [2], (note that
operators $\tilde{R}(p_{\mu})$ by definition commute with themselves
and with any other operator considered here),
%\begin{eqnarray}
$$ \tilde{f}(p_{\mu},p'_{\mu}) = \tilde{f}(p_{\mu})\tilde{f}(p'_{\mu}) +
\langle\tilde{a}^{+}(p_{\mu})\tilde{a}(p'_{\mu})\rangle
\langle\tilde{a}^{+}(p'_{\mu})\tilde{a}(p_{\mu})\rangle  +
\langle\tilde{a}^{+}(p_{\mu})\tilde{a}(p'_{\mu})\rangle
\tilde{R}^{+}(p'_{\mu})\tilde{R}(p_{\mu})
%+\\nonumber
%\\cr
+\langle\tilde{a}^{+}(p'_{\mu})\tilde{a}(p_{\mu})\rangle \tilde{R}^{+}(p_
{\mu})\tilde{R}(p'_{\mu}) ,$$
%\label{e49}
\begin{equation}
\tilde{f}(p_{\mu}) = \langle \tilde{a}^+(p_{\mu})\tilde{a}(p_{\mu})\rangle
 +\,|\tilde{R}(p_{\mu})|^2 .
\label{e50}
\end{equation}

This defines $D(p_{\mu},p'_{\mu}) = \tilde{f}(p_{\mu},p'_{\mu})/
[\tilde{f}(p_{\mu})\cdot\tilde{f}(p'_{\mu})] - 1$ in Eq. (\ref{e48}) in
terms of the operators $\tilde{a}(p_{\mu})$ and $\tilde{R}(p_{\mu})$
which in our case are equal to
%\begin{equation}
$$ \tilde{a}(p_{\mu}) =
\frac{\tilde{F}(p_{\mu})}{\omega - \tilde{K}(p_{\mu})}\quad {\rm and}\quad
\tilde{R}(p_{\mu}) =
\frac{\rho(\omega_{P},\epsilon)}{\omega - \tilde{K}(p_{\mu})} . $$
%\label{e51}
%\end{equation}
 This means, therefore,
that the correlation function $C_2(Q)$, as defined
by Eq. (\ref{e48}), is essentially given in terms of $\rho(\omega_{P},\epsilon)$
and the following two thermal averages for
the thermostat operators $F(\vec{p},t)$ (for details see [13]):
\begin{eqnarray}
\langle F^{+}(\vec{p},t)F(\vec{p}',t')\rangle =
\delta^3(\vec{p}-\vec{p}')\,\int \frac{d\omega}{2\pi}\,
               \left|\psi(p_{\mu})\right|^2\, n(\omega,\beta)e^{+i\omega(t-t')},
               \label{e52}\\
\langle F(\vec{p},t)F^{+}(\vec{p}',t')\rangle =
\delta^3(\vec{p}-\vec{p}')\,\int \frac{d\omega}{2\pi}\,
               \left|\psi(p_{\mu})\right|^2\, [1 + n(\omega,\beta)] e^{-i\omega(t-t')}.
               \label{e53}
\end{eqnarray}
%where $n(\omega) = \left\{\exp \left[(\omega - \mu)\beta\right] -
%1\right\}^{-1}$ is the number of (by assumption - only bosonic in our
%case) oscillators of energy $\omega$ in the reservoir
%characterized by parameters $\mu$ (chemical potential) and inverse
%temperature $\beta$ [13].
Notice that with only delta functions present
in Eqs.(\ref{e52}) and (\ref{e53}) we would have a situation in which the
hadronizing (or deconfined)
system would be described by some kind of {\it colored noise} only because
of the presence of $n(\omega,\beta)$ that carries the quantum properties . The
integrals multiplying these delta functions and depending on $(a)$ the momentum
characteristic of a heat bath $\psi(p_{\mu})$ and $(b)$ the assumed bosonic statistics
of produced secondaries resulting in factors $n(\omega,\beta)$ and $1+n(\omega,\beta)$,
respectively, bring the description of the system considered here closer to reality.
%It should be stressed at this point that, contrary to the majority of
%discussions of BEC \cite{BEC,Weiner}, we are working here directly in
%phase space, so far no space-time considerations
%were used.

We now see easily that the existence of BEC  (i.e., that $C_2(Q) >1$)
 is strictly connected with nonzero values
of the thermal averages Eqs. (\ref{e52}) and (\ref{e53}).
However, in the form
presented there, they differ from zero {\it only at one point},
namely for $Q=0$ (i.e., for $p_{\mu} = p'_{\mu}$). Actually, this is
the price we pay for the QFT assumptions tacitly made here, namely
for the {\it infinite} spatial extension and for the {\it uniformity}
of our reservoir. However, we know from the experiments in, e.g., [14,4-7] that
$C_2(Q)$ reaches its maximum at $Q=0$ and falls  towards its
asymptotic value of $C_2 = 1$ at large of $Q$ (actually  at $Q
\sim 1$ GeV/c). To reproduce the same behaviour by means of our
approach, we must replace the delta functions in Eqs.
(\ref{e52}) and (\ref{e53}) by functions with supports
larger than those limited to
one point only. This means that these functions should not be infinite
at $Q_{\mu} = p_{\mu}-p'_{\mu} = 0$ but remain more or less sharply
peaked at this point, otherwise remaining finite and falling to zero
at small but finite values of $|Q_{\mu}|$ (actually identical to
those at which $C_2(Q)$ reaches unity)
\begin{equation}
\delta(p_{\mu} - p'_{\mu})\, \Longrightarrow\, \Omega_0\cdot
\exp [-(p_{\mu}-p^{\prime}_{\mu})L^{\mu\nu}(r)(p_{\nu}-p^{\prime}_{\nu})].
%\sqrt{\Omega(q=Q\cdot r)}.
\label{e54}
\end{equation}
Here we  replace  the $\delta$-function with the smearing (smooth)
dimensionless generalized function
$\Omega(q=Q r)=\exp [-(p_{\mu}-p^{\prime}_{\mu})L^{\mu\nu}(r)(p_{\nu}-p^{\prime}_{\nu})]$,
where $L^{\mu\nu}$ is the structure tensor of the space-time size and it defines
the spherically-like domain of emitted (or produced) particles.
%$\Omega_0$ has the same dimension as the $\delta$ - function
%(actually, it is nothing else but $4$-dimensional volume restricting
%the space-time region of particle production).

We therefore tacitly introduce a new parameter, $r_{\mu}$, a $4$-vector that
 has the dimension of length. This defines the
region of {\it nonvanishing} particle density
with the space-time extension of the
particle emission source. Expression (\ref{e54}) must be understood
in the sense that $\Omega(Q r)$ is a function that,
 in the limit of $r\rightarrow \infty$,  {\it strictly} becomes  a
$\delta$ - function.
%Making such replacement in eq. (\ref{eq:theavc})
%one must also decide how to accordingly adjust $n(\omega)$ occurring
%there because now, in general, $\omega \neq \omega'$. In what follows
%we shall simply replace $n(\omega) \rightarrow n(\bar{\omega})$ with
%$\bar{\omega} = (\omega + \omega')/2$ (which, for classical particles
%would mean that $n(\omega) \rightarrow \sqrt{n(\omega)n(\omega')}$).
%In such way $r$ becomes new
%(and from the phenomenological point of
%view also the most important)
%parameter entering here together with
%the whole function $\Omega(Q\cdot r)$, to be deduced from comparison
%with experimental data.
With such a replacement, we now have
\begin{equation}
D(p_{\mu},p'_{\mu}) =
\frac{\sqrt{\tilde{\Omega}(q)}}{(1+\alpha)(1+\alpha')}
\cdot \left[ \sqrt{\tilde{\Omega}(q)} + 2\sqrt{\alpha \alpha'}
\right],  \label{e55}
\end{equation}
where
\begin{equation}
\tilde{\Omega}(q) = \gamma \cdot \Omega(q),~~
\gamma = \frac{n^2(\bar{\omega},\beta)}{n(\omega,\beta)\,n(\omega',\beta)}.
\label{e56}
\end{equation}
%with $\Lambda_{ch}$ being some characteristic
%scale of the deconfined state with the following properties:
%$\tilde{b}(k_{\mu})\rightarrow \tilde{a}(k_{\mu}),\, \tilde{R}(k_{\mu})=0$
%as $\Lambda_{ch}\rightarrow\infty $.
The coherence function $\alpha$ is
another very important one that summarizes our knowledge of
other than space-time characteristics of the particle emission source.
%In particular it contains
%the external static force $P$ present in the evolution equation
%(\ref{eq:Lang}). It is combined (in multiplicative way) with information
%on the momentum
%dependence of the reservoir (via $|\psi(k_{\mu})|^2$) and on the
%single particle distributions of the produced particles (via
%$n(\omega = \mu_T \cosh y)$ where $\mu_T$ and $y$ are, respectively,
%the transverse mass and rapidity).
Notice that $\alpha > 0$ only when $P \neq 0$. For $\alpha = 0$, we actually find
%\begin{equation}
$$ 1 < C_2(Q) < \chi(N)[1 + \gamma \Omega(Q r)](1+r_{f}\cdot Q + ...) , $$
%\label{e57}
%\end{equation}
i.e., it is contained between the limits corresponding to very large
(lower limit) and very small (upper limit) values of $P$.
Because of this, $\alpha$ plays the role of a {\it coherence}
parameter. Ignoring
the energy-momentum dependence of $\alpha$, and assuming that
$\alpha' = \alpha$, we get the expression
\begin{equation}
C_2(Q) \simeq \chi(N) \left \{1 + \lambda_{new}(m,\beta)\,e^{-q^2}
\left [1+\lambda_{corr}(m,\beta)\,e^{+q^{2}/2}\right ]
%\frac{2\alpha}{(1 + \alpha)^2}\cdot
%\sqrt{\tilde\Omega(q)}\, +\, \frac{1}{(1+\alpha)^2}\cdot
%\tilde\Omega(q)
\right \} (1+r_{f}\cdot Q + ...), \label{e58}
\end{equation}
where the new intercept function becomes as $\lambda_{new} =
\gamma(\omega,\beta)/(1+\alpha)^{2}$, and the new coherence
correction in the brackets of Eq. (\ref{e58}) carries an additional intercept function
$\lambda_{corr} = 2\,\alpha/\sqrt{\gamma(\omega,\beta)}$.
%which is formally {\it identical} with what has been obtained in
%\cite{Weiner} by means of QS approach.
%It has precisely the same
%form, consisting two $Q-$dependent terms containing the information
%on the shape of the source, one being the square of the other, each
%multiplied by some combination of the {\it chaoticity} parameter $p =
%1/(1+\alpha)$.
In fact, since $\alpha \neq \alpha'$
(because $\omega \neq \omega'$ and, therefore, the number
of states identified here with the number of particles with given
energy $n(\omega)$ is also different), we must use the
general form Eq. (\ref{e48}) for $C_2$ with details given by
Eqs. (\ref{e55}) and (\ref{e56}) and with $\alpha$ (\ref{e1}) depending on
the particle mass and such characteristics of the emission process as
the temperature $T$ and
chemical potential $\mu$ occurring in the definition of $n(\omega)$.
Note that Eq. (\ref{e58}) differs from the usual
empirical parameterization of $C_2(Q)$ [15,16,4-7],
\begin{equation}
C_2(Q) \sim (1 + \lambda \,e^{-q^2})(1+ aQ + ...)
, \label{e59}
\end{equation}
which is nothing other but the Goldhaber parameterization [17]
at $a=0$
%$$C_2(Q) = 1 + \lambda e^{-Q^2\,r^2}$$
with $0< \lambda <1$ being a free parameter adjusting the observed
value of $C_2(Q=0)$, customary called a "coherence strength factor"
or "chaoticity"  of  $\lambda =0$ for fully coherent sources and  $\lambda =1$
for fully incoherent sources; $a$ is a c-number, with
$\Omega(Q r)$ usually represented as Gaussian.
%Recently eq. (\ref{e59}) has found strong theoretical support expressed in
% detail in [15].
 %In \cite{Weiner} one introduces instead the notion of
%partially coherent fields representing produced particles, i.e., one
%has only one source, which produces partially coherent fields.
% Our
%approach is similar as we describe our particle by operator
%$b(\vec{k},t)$, which consists of two parts, cf. eq. (\ref{eq:apR}),
%one of which depends on the external static force $P$. The action of
%this force is to {\it order phases} of particles in our source
%(represented by the heat bath). The strength of this ordering depends
%on the value of the external force $P$. In any case, for $P \ne 0$,
%it demonstrates itself as a {\it partial coherence} \cite{FOOT-1}.
Returning  to the particle mass dependence of $C_2(Q)$ (the correlation radius,
in particular), we find that this dependence comes from
$\alpha$-coherence function (\ref{e1}) containing the operator kernel
$\tilde K(p_{\mu})$, defined correctly up to the second-order of the scalar
coupling constant $g^2$ in Eq. (\ref{e44}) within the framework of the QFT.

 The $\alpha$ - representation in (\ref{e1}) must be clarified:
In fact, $M^{4}_{ch}$ can be  broken down into two parts: $M^{4}_{ch}\rightarrow
M^{(0)}_{ch}\cdot \bar M^{3}_{ch}$, where $M^{(0)}_{ch}$ is the small massive
scale characterized by the time-like scale $r^0$, while $\bar M_{ch}$ is
the characteristic mass associated with the spatial inverse components $r_{z},\,r_{t}$,
 $\bar M^{3}_{ch}\sim (r_{z} r^{2}_{t})^{-1}$. Taking into account the
properties of the distribution $[\rho (\omega_{P},\epsilon)]^{2}$, we can
suggest the following replacement $M^{4}_{ch}[\rho (\omega_{P},\epsilon)]^{2}
\rightarrow [M^{(0)}_{ch}\rho (0,\epsilon)][\bar M^{3}_{ch}\,\rho (\omega_{P},\epsilon)]$,
where the first multiplier is of the order of $O(1)$, while the second reflects the
massive scale $\mu_{ch}$ of the particles production region, i.e.,
$\bar M^{3}_{ch}\,\rho (\omega_{P},\epsilon)\sim O(\mu_{ch}^2)$, $\mu_{ch} < m$. Thus,
\begin{equation}
 \alpha \sim O \left[\frac{\mu_{ch}^2}{{\vert\omega -
\tilde{K}(p_{\mu})\vert}^{2}\,n(\omega,\beta)}\right ] .
\label{e60}
\end{equation}

Let us return to the problem of the $Q$-dependence of the BEC. One more
remark is in order here: The problem with the
$\delta(p_{\mu}-p'_{\mu})$ function encountered in two particle
distributions does not exist in the single particle distributions
that are in our case given by Eq. (\ref{e50}), and which can be written as
%$\tilde{f}(k_{\mu}) \propto
%\langle \tilde{a}^+(k_{\mu})\tilde{a}(k_{\mu})\rangle\, +\,
%|\tilde{R}(k_{\mu})|^2 \, \sim (1+\alpha)\langle
%\tilde{a}^+(k_{\mu})\tilde{a}(k_{\mu})\rangle$.
%To be more precise
\begin{equation}
\tilde{f}(p_{\mu})\, = \, (1+\alpha) \cdot \Xi(p_{\mu},p_{\mu}) ,
 \label{e61}
\end{equation}
where $\Xi(p_{\mu},p_{\mu})$ is the one-particle distribution function
for the "free" (undistorted) operator $\tilde{a}(p_{\mu})$, namely
%\begin{equation}
$$ \Xi(p_{\mu},p_{\mu})\, =\langle\tilde{a}^+(p_{\mu})\tilde{a}(p_{\mu})\rangle
= \Omega_0 \cdot \left| \frac{\psi(p_{\mu})}
                           {\omega - \tilde{K}(p_{\mu})}\right| ^2
n(\omega,\beta) . $$
%\label{e62}
%\end{equation}
Notice that the actual shape of $\tilde{f}(p_{\mu})$ is dictated by both
 $n(\omega)$ (calculated for fixed temperature
$T$ and chemical potential $\mu$ at energy $\omega$, as given by
the Fourier transform of  field operator $\tilde{K}$  (\ref{e44}) and the
shape of the reservoir in the momentum space provided by
$\psi(p_{\mu})$), and by the $\delta$ - like distribution  of external
force $\rho (\omega_{P},\epsilon)$.
%They are both unknown, but because these details do not enter the BEC
%function $C_2(Q)$, we shall not pursue this problem further. What is
%important for us at the moment is that both the coherent and the
%incoherent part of the source have the same energy-momentum
%dependence (whereas in other approaches mentioned here they were
%usually assumed to be different).
 On the other hand, it is clear from Eq.
(\ref{e61}) that $\langle N\rangle = \langle N_{ch}\rangle +
\langle N_{coh}\rangle$, where $\langle N_{ch}\rangle$ and $\langle
N_{coh}\rangle$ denote multiplicities of particles produced
chaotically and coherently, respectively.

The $m$ - dependence of $C_{2}(Q,m)$ is essentially given by
\begin{equation}
\alpha (m^{2},\beta) \sim \frac{n(\omega,\beta)}{\bar{m}^{2}(\omega,\beta)\,L^{2}_{ch}},
 \label{e65}
\end{equation}
which is nothing other but the effective number of Bose-particles in the plane
phase-space with the size $L_{ch}$ having the mean mass $\bar{m}(\omega,\beta)=
m\,n(\omega,\beta)$. Obviously, $\alpha\rightarrow 0$ as
$n(\omega,\beta)\rightarrow\infty$ (Goldhaber parameterization) and
$\alpha\rightarrow\infty$ as $n(\omega,\beta)\rightarrow 0$ (that is trivial
result with $C_{2}(Q)\simeq 1$). Notice, that the condition $\alpha (m,\beta) <1$
always yields at sufficiently strong temperature $T$, and its validation would be
very weak when $T\rightarrow 0$. In the latter case the heat bath is simply absent.

On the other hand, neglecting the energy-momentum dependence of $\alpha\, (\alpha^{\prime})$,
the latter can be estimated within the formula
$$\alpha\simeq\frac{2-\bar C_{2}(0) + \sqrt{2-\bar C_{2}(0)}}{\bar C_{2}(0)-1}, $$
where $\bar C_{2}(0) = C_{2}(0)/\chi (N)$, and we assume that the BEC function
at $Q=0$, $C_{2}(0)$, and the mean multiplicity
$\langle N\rangle$ are known from the experiment.
Within the formula (\ref{e65}) and the Eq. (\ref{e58}), it is evidently that an
increasing of the particle mass $m$ leads to an enhancement of the intercept
function $\lambda_{new}(m,\beta)$. Therefore in the case of the investigation of the correlations
between heavy particles the chaotic coherence effects become negligible $(\alpha\rightarrow 0)$.

Finally, we present the lower bound on the particle emitting source size:
%\begin{equation}
$$ r_{ch}\sim L_{ch} > \frac{1}{m\,\sqrt{n(\beta)}}
{\left [\frac{1}{{\gamma(n)}/{\Delta \bar C^{max}_{2}} -1}\right ]}^{\frac{1}{4}}, $$
% \label{e63}
%\end{equation}
where $\Delta \bar C^{max}_{2} = \bar C^{max}_{2} - 1$, $\bar C^{max}_{2} = C^{max}_{2}/\chi (N)$,
 and $C^{max}_{2}$ is the maximal value of the $C_{2}$-function in the vicinity of $Q=0$.

The lower bound on the particle emitting source size
carries the dependence of: \\
%- $\alpha$ and $\alpha^{\prime}$ strength couplings of coherence;\\
- the particle mass $m$ given by $\alpha$ and $\alpha^{\prime}$ which are
defined within
(\ref{e60}) with the kernel operator $\tilde K(p_{\mu})$ (\ref{e44}) calculated in the QFT;\\
- the mean multiplicity  factor $\chi(N)$ defined within formula (\ref{e48});\\
- the maximal value of the BEC function $C_{2}$;\\
- the absolute temperature of a heat bath
and the chemical potential (the $\gamma (n)$-factor defined in Eq. (\ref{e56})).

%{\bf 6.}
%{\centerline {\bf {6. Conclusion }}}
\section { Conclusion }

To summarize: using the QFT, supplemented by Langevin-like evolution Eqs.
(\ref{e29}) and  (\ref{e30}) to describe hadronization (or deconfinement)
processes, we have derived the two-particle BEC function in a form explicitly showing the origin of
both the so-called coherence (and how it influences the structure of the
BEC) and the $Q$-dependence of the BEC represented by the correlation
function $C_2(Q)$. The dynamic source of coherence is identified in
our case with the existence of a constant external term $P$ in the
evolution equation.
%Its influence turns out to be identical with the
%one obtained before in the QS approach \cite{Weiner} and is described
%by eq. (\ref{eq:res}).
%Its action is to order phases of the produced
%secondaries.
 Therefore, for $P\rightarrow \infty$, we have all phases
aligned in the same way, and $C_2(Q) =1$. This is because coherence
has already been introduced on the
level of a particle production source as a property of the fields
or operators describing produced particles.
%Dividing instead the
%hadronizing source itself into coherent and chaotic subsources leads
%to results obtained in \cite{ALS} and given by eq. (\ref{eq:usual}).
%The controversy between results given by \cite{Weiner} and \cite{ALS}
%is therefore explained: both approaches are right, one should only
%remember that they use different descriptions of the notion of
%coherence.
It is therefore up to the experimenter to decide which
proposition is followed by nature: the simpler formula
(\ref{e59}) or the rather  more complicated Eq. (\ref{e48}) in combination
with  Eq. (\ref{e55}).
%From Fig. 1 one can see that
%differences between both forms are clearly visible, especially for
%larger values of the coherence function $\alpha$.
%\begin{figure}
%\noindent
%\centerline{\epsfig{c2.eps, width =72mm}}
%\caption{Shapes of $C_2(Q)$ as given by eq. (\protect\ref{eq:C2final})
%for different choices of smearing functions $\Omega(q)$ (cf. text for
%details).}
%\label{fig:plot1}
%\end{figure}
%\begin{figure}
%\noindent
%\centerline{\epsfig{file= alphabw.eps, width= 72mm}}
%\caption{Shapes of $C_2(Q)$ as given by eq. (\protect\ref{e58}) -
%upper panel and for the truncated version of (\protect\ref{e58})
%(without the middle term) - lower panel. Gaussian shape of $\Omega(q)$ was
%used in both the cases.}
%\label{fig:plot2}
%\end{figure}

With our approach, it is also clear that the form of $C_2$
reflects the distributions of the space-time separation between the
two observed particles.

Finally, we would like to stress that our discussion is so far
limited to only a single type of secondaries  produced.
%It is
%also aimed at a description of hadronization or deconfinement understood
%as kinetic freeze-out in some more detailed approaches.
%So far we were not
%interested in the other (highly model dependent) details of the
%particle production process.
This is enough to attain our general
goals, i.e., to explain the possible dynamic origin of coherence in
BEC, the origin of the specific shape of the correlation $C_2(Q)$
functions, and to explain the dependence of the correlation radius on the
particle mass due to  coherence function $\alpha$,
as seen from the the QFT perspective. Actually, $r$ decreases with the mass $m$.
It is then plausible that, in describing the general BEC effect,
they should be combined somehow, especially if the experimental data
indicate such a need.

The final note concerns to that the ZEUS at HERA resulted that $r$ values for
pions and kaons are not so different and the effects comes from heavier particles.
Therefore more precise measurements from different processes are necessary.

%Part of this work is based on a collaboration with G. Wilk and O. Utyuzh.
%I have greatly benefited from our many animated discussions. I am also grateful
%to S. Tokar for his kind hospitality and for many fruitful discussions.
%But to do so our approach should first
%be generalized to allowing for production
%of different types of secondaries and allow also for resonance production
%and final state interactions (both of strong and Coulomb origin). \\


\begin{references}
\bibitem{13}

The origin of the two parameters occurring at this point,
                 the temperature $\beta =1/T$ and chemical potential $\mu$
                 is the Kubo-Martin-Schwinger condition that
                 $\langle a(\vec{p}',t')a^{+}(\vec{p},t)\rangle =
                 \langle a^{+}(\vec{p},t)a({p}',t-i\beta)\rangle
                 \cdot \exp(-\beta \mu)$, see [2].
\end{references}
\end{document}